\documentstyle[12pt]{article}
\begin{document}

\begin{titlepage}
\rightline{June 2013}
\vskip 2.2cm

\centerline{\large \bf 
Thin disk of co-rotating dwarfs: 
}
\vskip 0.4cm
\centerline{\large \bf 
a fingerprint of dissipative (mirror) dark 
matter?
}

\vskip 1.6cm
\centerline{R.~Foot$^{a}$ and Z.~K.~Silagadze$^{b}$~\footnote{
E-mail address: rfoot@unimelb.edu.au, \hspace{1mm}
silagadze@inp.nsk.su}
}

\vskip 0.5cm
\centerline{\it $^a$ School of Physics,}
\centerline{\it University of Melbourne,}
\centerline{\it Victoria 3010 Australia}
\vskip 0.5cm
\centerline{\it $^b$ Budker Institute of Nuclear Physics SB RAS and}
\centerline{\it Novosibirsk State University,}
\centerline{\it  630 090, Novosibirsk, Russia}
\vskip 2cm
\noindent
Recent observations indicate that about half of the dwarf satellite galaxies 
around M31 orbit in a thin
plane approximately aligned with the Milky Way.
It has been argued that 
this observation along with several other features can be explained
if these dwarf satellite galaxies originated as tidal dwarf galaxies formed 
during an ancient merger event.
However if dark matter is collisionless then tidal dwarf galaxies should be 
free of dark matter - 
a condition that is difficult to reconcile with observations
indicating that dwarf satellite galaxies are dark matter dominated.
We argue that dissipative dark matter candidates, such as mirror dark matter,
offer a simple solution to this puzzle.
\end{titlepage}

Recently the discovery of a vast, thin plane of co-rotating dwarf galaxies 
orbiting the Andromeda galaxy \cite{11} has been reported.
It was found that about half of the satellites of the Andromeda M31 
galaxy belong to a vast extremely thin planar structure (about 400~kpc 
in diameter but only about 14~kpc in thickness). The structure
is co-planar with the Milky Way to M31 position vector and is almost 
perfectly aligned with the pole of the Milky Way's disk.
Of the 15 satellites in this thin plane, 13 of these were 
found to be co-rotating 
sharing the same direction of angular momentum.  A similar planar structure 
but not so extremely thin and not so numerous
(nine satellites with one counter-orbiting)
was found earlier 
among Milky Way's dwarf satellites \cite{12}.  Such 
vast coherent planar structures constitutes an interesting puzzle from the 
point of view of
the  standard Concordance Cosmology \cite{13}.

Such structures can be understood if the satellite galaxies originated
as tidal dwarf galaxies.
It was suggested by Zwicky long ago that 
new dwarf galaxies can be formed as a result of violent disruptions of 
galaxies during close encounters \cite{14}. When two galaxies interact,
tidal forces rip out matter from the galactic discs thus providing enough 
intergalactic tidal debris from which new dwarf galaxies (TDGs -- Tidal 
Dwarf Galaxies) are formed. In such a tidal scenario planar distribution of
newly formed TDGs as well as their correlated orbital moments are a natural
outcome as the TDGs are formed from a common tidal tail in a plane defined 
by the orbit of the interaction \cite{12}.
Interestingly, it was shown that significant amount of counter-orbiting 
material emerges quite naturally in tidal interactions of disc galaxies
\cite{15}. Therefore, a tidal scenario can explain even the existence 
of the satellites orbiting counter to the bulk of other dwarf galaxies
in the planar structure.

It was argued \cite{16} that the kinematical and morphological properties
of M31 can be explained by assuming a single major merger event at the M31 
location about 8.7 Gyr ago during which TDGs have been formed. Further 
studies revealed that the Milky Way disk of dwarf satellites, the so called 
Vast Polar Structure, also can be explained through the tidal tail of this 
ancient merger event \cite{17}.
The Vast Thin Disk of Satellites (VTDS) around M31 could have been predicted 
before its discovery by Ibata et al. \cite{11} (see also \cite{18}). Indeed, 
it was found post factum that the induced tidal tails by the above mentioned 
merger event are lying just in the VTDS plane and the positions and 
velocities of the VTDS dwarfs are quite accurately reproduced without any 
need for fine tuning the model \cite{19}.
This model also
seems to provide a reasonable framework to 
understand some other puzzling features in the Local Group \cite{19}.

If dark matter is collisionless, then there is a problem with this simple 
picture, called the Zwicky Paradox in \cite{20}: this scenario assumes that 
the Milky Way and Andromeda dwarf satellites are tidal dwarf galaxies. Such 
galaxies should be free of nonbaryonic dark matter if dark matter is 
collisionless \cite{21}. Detailed observations indicate on the contrary that 
they are dark matter dominated \cite{22}\footnote{
The Zwicky Paradox will be eliminated if the satellite galaxies are out of 
virial equilibrium due to tidal perturbations upon close perigalactic 
passages (see \cite{23} and references therein). Such explanation of the 
apparent dark matter content of dwarf spheroidals might be plausible for
the closest to the Milky Way dwarfs with elongated stellar structures but
cannot explain the apparent dark matter content of the entire satellite
population \cite{22}. In particular, recent observations of the two distant
dwarf satellites in the outskirts of the M31 system indicate that they are 
also dark matter dominated \cite{24}.}.    
 


The reason why TDGs are expected to be devoid of dark matter is the 
following \cite{21,26}. TDGs, being recycled galaxies formed from the 
collisional debris, are gas dominated and their formation is dissipation
supported under the crucial condition that the progenitor disks contain
sufficiently massive and extended gas components \cite{28}. Standard WIMP
dark matter is assumed to be dissipationless. As a result such dark 
matter surrounds galaxies in the form of large nearly spherical  halos
supported by random motions. On the contrary, gas particles in the rotating
disks of progenitor colliding galaxies have nearly circular coplanar orbits.
This difference in phase space distributions indicates that galactic 
collisions effectively segregate dark and baryonic matter: tidal tails 
from which TDGs are subsequently formed  
should contain very little non-dissipative dark matter \cite{26,28}. 


If dark matter is dissipative then the formation of a dark disk in parallel 
to the ordinary baryonic disk is possible in galaxies.  
Hidden sector dark matter, where dark matter resides in a hidden sector 
with its own gauge group, $G'$,
can be dissipative if $G'$ contains an unbroken $U(1)'$ factor (sometimes 
called `dark photon' in the literature).
This means that  the Lagrangian describing the particle physics decomposes 
into two sectors, one describing the standard particles and forces, and another
which will contain the dark matter (we neglect possible interactions other 
than gravity between two sectors for a moment):
\begin{eqnarray}
{\cal L} = {\cal L}_{SM} + {\cal L}_{dark}
\ .
\end{eqnarray}
Mirror dark matter, where the hidden sector is exactly isomorphic to the 
ordinary sector \cite{39} (see also \cite{40,41,42}), 
is an interesting special case 
of such a theory, for reviews see e.g. \cite{43,44,45,46,48} and
references there-in for a more complete bibliography.
In this case there is an unbroken symmetry, which can be interpreted as 
space-time parity, which maps 
each ordinary particle, $e, \nu, u, d, ..., \gamma,...$ onto a
mass degenerate mirror partner, which we denote with a prime ($'$): 
$e', \nu', u', d', ..., \gamma',...$.
The symmetry ensures that the gauge self-interactions (mirror electromagnetism 
etc) have the same form and strength in the mirror sector as they do in the 
ordinary sector.
In addition to gravity, the ordinary and mirror sector particles can interact 
with each other via 
the kinetic mixing interaction, which is both 
gauge invariant and renormalizable \cite{he,holdom}:
\begin{eqnarray}
{\cal L}_{mix} = \frac{\epsilon}{2} F^{\mu \nu} F'_{\mu \nu}
\label{kine}
\end{eqnarray}
where $F_{\mu \nu}$ ($F'_{\mu \nu}$) is the field strength tensor for the 
photon (mirror-photon).

In the mirror dark matter framework, spiral galaxies like
M31 and the Milky Way, are currently composed of mirror particles
predominately in a (roughly) spherical, pressure supported halo \cite{sph}. 
Additionally there may be remnant old mirror stellar objects (mirror white 
dwarfs, mirror neutron stars etc) -- the spherical halo in spirals is too hot 
for significant mirror star formation to occur at the present epoch.
The spherical halo dissipates energy due to bremsstrahlung and other 
processes. This energy can be replaced by the energy produced from ordinary 
type II supernova. This requires kinetic mixing
of strength $\epsilon \sim 10^{-9}$ (to produce the energy via plasmon decay 
processes in the core of type II supernova \cite{raffelt,sil}) and also that 
the mirror particle halo has a sufficient mirror
metal component - at least around one percent by mass (so that the 
plasma can absorb this energy via photoionization) \cite{sph,50}.
The balancing of dissipated energy with energy produced from ordinary 
supernovae might be enforced by the dynamics. This energy balance condition 
has been used in ref.\cite{50} to derive two scaling relations governing the 
(current) dark matter density profile in spiral galaxies,
both of which are in agreement with observations \cite{salucci}.
Ref.\cite{50} also pointed out that stringent constraints on self-interacting 
dark matter from elliptical galaxies and the Bullet Cluster (see also 
\cite{zurab}), can potentially be evaded.


Although mirror dark matter seems to provide a successful picture to describe 
the current properties of galaxies, the role of its dissipative nature on
the galaxy formation and early history is a topic that has been neglected.  
Naturally any discussion in this direction is speculative and certainly 
preliminary. With this note of caution, we now proceed to outline the 
emerging picture.

To understand the early growth of structure, one has to go back to early times, when density perturbations were small, i.e. $\delta \rho/\rho \ll 1$,
known as the linear regime. BBN and CMB observations require
asymmetric initial conditions,
$T' \ll T$, \ $\Omega_{b'} \approx 5\Omega_b$, to hold.
With $T' \ll T$, mirror-hydrogen recombination occurs very early, much earlier than ordinary hydrogen
recombination. Prior to hydrogen (mirror-hydrogen) recombination, one has a ordinary (mirror) plasma strongly coupled to photons (mirror-photons).
Perturbations (in Fourrier space) which enter the horizon prior to recombination undergo acoustic oscillations which suppresses the growth
of structure. Since mirror-hydrogen recombination occurs much earlier than hydrogen recombination 
this suppression is much less effective for mirror baryons.
For this reason the growth of mirror baryon density perturbations are well in advance of the ordinary baryonic density perturbations
by the end of the linear regime\cite{berlss,rrvlss}.

Eventually the perturbations reach the point where $\delta \rho/\rho \sim 1$ and gravitational collapse can occur,
and linear perturbation theory is no longer valid.
To gain some insight, imagine first
a uniform collapsing spherical density perturbation of mass over-density $\rho_0$ and
temperature $T$.  For simplicity consider a collapsing plasma composed of fully ionized mirror-helium,
so that  $n_{e'} \approx 2n_{He'} \approx  2n_T/3$, where $n_T$ is the total particle number density.
The cooling rate per unit volume due to thermal bremsstrahlung is:\footnote{
The equations have been adapted from the usual treatment of baryonic physics, reviewed in e.g. \cite{book}.}
\begin{eqnarray}
\Gamma_{cool} = n_{e'}^2 \Lambda
\end{eqnarray}
where $\Lambda \sim 10^{-23}\ {\rm erg}\ {\rm cm^3}\ {\rm s^{-1}}$ for $T \sim 100$ eV.
The cooling time scale, $t_{cool}$, is defined from $\Gamma_{cool} t_{cool} \approx n_T (3/2) T$,
i.e.
\begin{eqnarray}
t_{cool} &\approx & {9T \over 4\Lambda n_{e'}} \nonumber \\
& \sim & 100 \ \left({T \over 100 \ {\rm eV}}\right) \left( {10^{-2}\ {\rm cm^{-3}} \over n_{e'}}\right) \ {\rm Myr} 
\ .
\end{eqnarray}
Another important time scale is the
free fall time, $t_{ff}$, for the uniform collapsing spherical density perturbation.
This time scale is given by:
\begin{eqnarray}
t_{ff} = \sqrt{{3\pi \over 32 G_N \rho_0}}\ 
\end{eqnarray}
where $\rho_0 = n_{He'} m_{He'} = n_{e'} m_{He'}/2$.
In the absence of any heat source, 
perturbations satisfying $t_{cool} < t_{ff}$ can collapse unimpeded.
Evaluating, $t_{cool}/t_{ff}$, we have:
\begin{eqnarray}
{t_{cool} \over t_{ff}} \sim 0.3 \ \left( {T \over 100 \ {\rm eV}}\right)\left( {10^{-2}\ {\rm cm^{-3}} \over n_{e'}}\right)^{1/2}
\ .
\end{eqnarray}
Evidently there is no impediment for collapse of (typical) spiral galaxy sized perturbations due to pressure effects.
We thus have that in the early period (linear regime) mirror baryonic perturbations grow faster than 
baryonic ones given the necessary initial condition $T' \ll T$, while in the nonlinear regime, the
pressure does not impede collapse of mirror baryonic structures. Therefore mirror baryons form structures first.

Within these structures the mirror baryons can collapse forming mirror stars, either in the free-fall phase
or later in a dark disk.
Either way rapid mirror star evolution is envisaged \cite{star}, potentially producing also mirror supernovae.
With $\epsilon \sim 10^{-9}$ mirror
supernovae would provide a huge flux of ordinary (x-ray?) photons. It is very natural to suppose that this
radiation might have been responsible for the re-ionization of ordinary matter 
at high redshift $z > 6$ inferred to exist from CMB observations.
However the plasma cannot efficiently absorb radiation once the ordinary matter is ionized;
the Thomson scattering cross-section is too small and photoionization is ineffective due to the
low metal content at this early time.
The ordinary baryons, therefore, should ultimately collapse potentially forming a separate disk. 
Gravitational interactions between the baryonic disk and a mirror baryonic disk (assuming both form) would cause them 
to merge on a fairly short time scale \cite{30}.

Eventually ordinary star formation and hence also ordinary supernovae will occur in the merged disk.
With $\epsilon \sim 10^{-9}$ these ordinary supernovae would provide a huge source of  mirror-photons,
heating the mirror plasma component via photoionization (given 
the mirror metal enrichment of the plasma by this time).
With sufficient heating,
the mirror gas component would expand to form the spherical halo, where 
ultimately the energy supplied by ordinary supernova heating balances the 
energy dissipated due to bremsstrahlung (discussed earlier). The end result is that today, we might expect that the
galactic disk would contain just the ordinary baryons and a remnant mirror 
stellar component, which would be surrounded by a (roughly) spherical mirror particle halo. 
However, the epoch of TDG formation (roughly 8.7 Gyr ago in the model
of ref.\cite{16}) might have been early enough for the mass density
of the disk to have been dominated by the mirror gas component.
If this is indeed the case then TDGs can form from this dissipative material. 
These TDG can further accrete dark matter
over time as they travel through the outer halo of their host galaxy (M31 
or Milky Way). Thus, mirror dark matter, and perhaps closely related hidden 
sector dark matter models, might thereby explain why dwarf satellite galaxies 
are dark matter dominated. 




\vskip 0.5cm
\noindent
{\large \bf Acknowledgments}
\vskip 0.3cm
\noindent The work of Z.K.S. is supported by the Ministry of Education and
Science of the Russian Federation and in part by Russian Federation President
Grant for the support of scientific schools NSh-5320.2012.2  and by
RFBR grant 13-02-00418-a. The work of R.F. was supported by the Australian 
Research Council.

\end{document}